\documentclass[aps,showpacs,onecolumn,floatfix,amsmath,amssymb,nofootinbib]{revtex4-1}
\usepackage{float}
\usepackage{epsfig}
\usepackage{graphicx}
\usepackage{subfigure}
\usepackage{color}
\usepackage{hyperref}
\begin{document}

\title{Thermodynamics of the Euler-Heisenberg-AdS black hole}
\author{Daniela Magos$^{1}$}
\email{dmagos@fis.cinvestav.mx}

\author{Nora Breton$^{1}$}
\email{nora@fis.cinvestav.mx}

\affiliation{$^{1}$Departamento de F\'isica, Centro de Investigaci\'on y de Estudios
Avanzados
del I.P.N.,\\ Apdo. 14-740, Mexico City, Mexico.\\
}

\begin{abstract} 
We generalize the solution to the Euler-Heisenberg (EH) theory coupled to gravity that represents
a nonlinearly charged static black hole (BH) introducing the cosmological constant;
the obtained solution is characterized by four
parameters: mass $M$, electric charge $Q$, cosmological constant $\Lambda$ (positive
or negative) and the Euler-Heisenberg theory parameter $a$. Then we briefly analyze some BH features like
horizons, electromagnetic field and geodesics. We mainly focus on its
thermodynamic properties in the extended space where the  anti--de Sitter
parameter is interpreted as the pressure; we show the consistency between the Smarr
formula and the first law of BH thermodynamics, interpreting the parameter of the EH theory as 
the vacuum polarization. We determine the equation of state and the
critical points; the critical volume determines two branches of BHs, one near
Maxwell behavior and a second one manifestly nonlinear electromagnetic. Moreover,
the analysis of the Gibbs free energy indicates that two phase transitions can  occur; 
we also construct the coexistence curve $P$-$T$ where the different phases of the BH can be observed; the critical point is
characterized by the standard mean field theory exponents, and the critical variables satisfy $P_{\rm crit} v_{\rm crit}/ T_{\rm crit} = 3/8 $ plus the terms of the ten thousandths order in the Euler-Heisenberg parameter $a$.
\end{abstract}


\maketitle

\section{Introduction}

Black Hole (BH) thermodynamics gained interest related to phase transitions after the seminal work
by Hawking and Page \cite{Hawking-Page} where they showed that the Schwarzschild BH
in anti--de Sitter (AdS) spacetime suffers a phase transition from a Schwarzschild BH
metric to a thermal AdS metric.
Later on it was shown that charged AdS BHs present 
a first order {\it small/large} BH phase transition,
in a canonical ensemble, resembling nonideal fluid phase transitions
\cite{Chamblin}; for the charged Reissner-Nordstrom anti--de Sitter (RN-AdS)
BH there exists a small-large BH transition, in analogy with the gas-liquid 
phase transition of the van der Waals fluid.
Moreover, the interpretation of the
cosmological constant as the thermodynamic pressure and of the volume as its thermodynamic
conjugate variable leads  to consistency between the Smarr relation, the formula
for the mass of the BH in terms of the rest of BH parameters, and the first law of
BH thermodynamics, from which Smarr's formula can be derived from scaling arguments \cite{Kubiznak2017}.

Another aspect of interest of the anti--de Sitter spacetimes is related to the holographic
correspondence between gravity systems and the conformal field theory, the AdS/CFT duality;
then  systems in AdS spacetimes are worthwhile to study since they admit a gauge duality
description through a thermal field theory \cite{Witten98}.  

Since there are critical phenomena for the charged-AdS BHs, 
it is important to explore phase transitions  of these objects in the context of 
nonlinear electrodynamics (NLED).
In \cite{Mann} the study of the Born-Infeld BH is presented,
the solution to the Einstein equations coupled to the nonlinear electrodynamics of 
Born-Infeld (BI). To study the BI-BH thermodynamics the authors defined a new
quantity, the BI vacuum polarization, the thermodynamic  conjugate variable of the BI parameter, 
and they showed that this quantity is required for consistency
between the first law of thermodynamics and the corresponding Smarr relation.
Inspired by this work we explore the Euler-Heisenberg black hole, the solution to 
the Einstein gravity coupled to the NLED proposed by Euler 
and Heisenberg (EH) in 1936 \cite{EH}, derived directly from quantum electrodynamics (QED)
to one-loop approximation. This theory is well established and  the nonlinear electromagnetic effects that arise at the critical or Schwinger fields, $E_c$,  of the order of  $10^{18}$ V/m for an electric field  (or $10^9$ Tesla for a magnetic field ) are currently being experimentally tested \cite{bordin}.

In this paper we address the thermodynamic study of the 
Einstein-Euler-Heisenberg (EEH) black hole in the
extended phase space,
where the cosmological constant is considered the dynamical pressure and the
corresponding conjugate quantity is the
volume. In order to consider the thermodynamics of the EEH BH in the extended space,
we first generate the EEH solution with cosmological constant, $\Lambda$,
that is interpreted as the thermodynamic pressure;
then we show the consistency between the first BH thermodynamics law and
the Smarr's relation for this system. The introduction of $\Lambda$ allows the
existence of circular stable trajectories of massless test particles in 
the neighborhood of the EEH BH. We then focus on the thermodynamic 
phase transitions of the BH, and we derive the equation of state, as well as the critical 
quantities. By means of the analysis of the Gibbs potential we determine the critical points and the fulfilment of the Maxwell's area law.  The near critical state renders the 
relation between critical exponents, whose value is the same than in standard mean field theory. Moreover, the critical  variables fulfil  $P_{\rm crit} v_{\rm crit}/ T_{\rm crit} = 3/8 $  plus terms of the ten thousandths order in the Euler-Heisenberg parameter $a$.

In the next section we briefly give the basics on Euler-Heisenberg (EH) nonlinear
electrodynamics.  In Sec. III we derive the generalization of the EEH-BH introducing 
the cosmological constant, $\Lambda$, that in principle can be positive or negative. 
By introducing the cosmological constant to the already known EEH
black hole  \cite{Yajima2001}, \cite{Ruffini2013}, \cite{Amaro},
the parameters that characterize the EEH-$\Lambda$ BH are $(M, Q, a,
\Lambda)$. In the same section effects on geodesics arising from the introduction of $\Lambda$ 
are deduced from the analysis of the effective potential for a test particle.
In Sec. IV we restrict to $\Lambda < 0$ and the EEH-AdS BH thermodynamics is addressed. We derive the Smarr formula with the
thermodynamic variables in the extended space and show its consistency with the
first BH- thermodynamics law. From the behavior of the critical temperature versus
the EH parameter we found two branches of the solution: the Maxwell and the EH
branch, separated by a critical point. We present the equation of state and the
$P$-$V$ diagrams. The behavior of the specific heat at constant charge points to a phase transition that we confirm from the Gibbs potential that is derived and plotted as a function of the temperature as well as the coexistence $P$-$T$ curve. In this same section we analyze the critical exponents.
Conclusions are settled in the last section.
\section{Euler-Heisenberg theory}

In this section we give an overview of the EH theory coupled to gravity.
The four-dimensional action of general relativity with cosmological constant
$\Lambda$ coupled to nonlinear electrodynamics (NLED) \cite{Plebanski},
\cite{Salazar} is given by 

\begin{equation}
S=\frac{1}{4\pi}\int_{M^4} d^4x \sqrt{-g}\left[\frac{1}{4}(R-2\Lambda)-
\mathcal{L}(F,G) \right],\label{action}
\end{equation} 
where $g$ is the determinant of the metric tensor, $R$ is the Ricci scalar,
$\mathcal{L}(F,G)$ is the NLED Lagrangian that depends on the electromagnetic
invariants,    $F=\frac{1}{4}F_{\mu \nu} F^{\mu \nu}$ and $G=\frac{1}{4}F_{\mu \nu}
{^*F^{\mu \nu}}$ with $F_{\mu\nu}$ denoting the electromagnetic field strength
tensor and ${^*F^{\mu \nu}}=\epsilon _{\mu\nu\sigma\rho} F^{\sigma\rho}
/(2\sqrt{-g})$ its dual. The completely antisymmetric  tensor
$\epsilon_{\mu\nu\sigma\rho}$, satisfies
$\epsilon_{\mu\nu\sigma\rho}\epsilon^{\mu\nu\sigma\rho}=-4!$\\
The Lagrangian density for the Euler-Heisenberg NLED \cite{EH} is 
\begin{equation}
\mathcal{L}(F,G)=-F+\frac{a}{2}F^2+ \frac{7a}{8} G^2,\label{Lagrangian}
\end{equation}
where $a=8 \alpha^2/45 m^4$ is the Euler-Heisenberg parameter that regulates the intensity of the
NLED contribution; $\alpha$ is
the fine structure constant and $m$ is the electron mass (we take $c=1= \hbar$), such that the EH parameter
is of the order of $\alpha / E_c^2$.
For $a=0$ we recover the Maxwell electrodynamics $\mathcal{L}(F)=-F$.

Regarding NLED
there are two possible frameworks, one of them is the usual one ($F$ framework) in terms of
the electromagnetic field tensor $F^{\mu \nu}$.
Alternatively, there is the $P$ framework with the tensor $P_{\mu\nu}$ as the main field,
defined by
\begin{equation} 
P_{\mu\nu}= -(\mathcal{L}_F F_{\mu\nu}+ {^*F}_{\mu\nu} \mathcal{L}_G ),
\end{equation}
where the subscript $X$ in $\mathcal{L}$ denotes the derivative, $  \mathcal{L}_{X}= d
\mathcal{L} /d X$. In  the Euler-Heisenberg theory, $P_{\mu \nu}$ takes the form
\begin{equation}
P_{\mu\nu}=(1-a F)F_{\mu\nu} - {^*F}_{\mu\nu} \frac{7a}{4}G .\label{Pmunu_a1}
\end{equation}
This tensor corresponds to the electric induction {\bf D} and the
magnetic field {\bf H} and  Eqs. (\ref{Pmunu_a1}) are the constitutive relations
between {\bf D},  {\bf H} and the magnetic intensity {\bf B} and the electric field {\bf E} in the EH NLED. \\

The two independent invariants $P$ and $O$ associated to the P framework are defined as
\begin{equation}
   P=-\frac{1}{4} P_{\mu\nu} P^{\mu\nu}, \hspace{1cm}
   O= -\frac{1}{4} P_{\mu\nu} {^*P^{\mu\nu}},
\end{equation}
with $^*P_{\mu\nu}=\frac{1}{2\sqrt{-g}}\epsilon_{\mu\nu\rho\sigma}P^{\sigma\rho}$.
The Legendre transformation of $\mathcal{L}$ defines the Hamiltonian or structural
function $\mathcal{H}$,
\begin{equation}
\mathcal{H} (P,O)= -\frac{1}{2}P^{\mu\nu} F_{\mu\nu}-\mathcal{L}.
\end{equation}
Neglecting  the second and higher order terms in $a$, the structural function for the EH
theory takes the form \cite{Ruffini2013},
\begin{equation}
\mathcal{H}(P,O)= P-\frac{a}{2}P^2-\frac{7a}{8}O^2. \label{HamiltonianEH}
\end{equation}

The field equations are \cite{Salazar}
\begin{equation}
\nabla_\mu P^{\mu\nu}=0, \hspace{.7cm} G_{\mu\nu}+\Lambda g_{\mu\nu} =8\pi
T_{\mu\nu}.\label{motion}
\end{equation}
The energy momentum tensor $T_{\mu\nu}$ for the EH theory in the $P$ framework is
given by
\begin{equation}
T_{\mu\nu}=\frac{1}{4\pi}\left[(1-a P)P_\mu^\beta
P_{\nu\beta}+g_{\mu\nu}\left(P-\frac{3}{2}a
P^2-\frac{7a}{8}O^2\right)\right].\label{emtensor}
\end{equation}

In the next section we present the static spherically symmetric solution of
the EEH equations with cosmological constant.

\section{Electrically charged EEH-$\Lambda$ BH solution}

We determine the solution to the field Eqs. (\ref{motion}) for a static, spherically
symmetric metric of the form
\begin{equation}
ds^2= -f(r)dt^2 + f(r)^{-1}dr^2 + r^2(d\theta^2 +\sin^2{\theta} d\phi^2),\label{metric}
\end{equation}
with $f(r)=1-\frac{2m(r)}{r}$. Regarding the electromagnetic field, restricting to
an electric charge $Q$, the symmetry of the spacetime allows the nonvanishing
components, 
\begin{equation}
P_{\mu\nu}= \frac{Q}{r^2}\delta^0_{[\mu}\delta^1_{\nu]}; \label{Pmunu}
\end{equation}
then the electromagnetic invariants are
\begin{equation}
P=\frac{Q^2}{2r^4},\hspace{.7cm} O=0  \label{PO} . 
\end{equation}
Substituting these into the (0,0) component of the field Eqs. (\ref{motion}),
we find
\begin{equation}
\frac{dm}{dr} = \frac{Q^2}{2r^2}-\frac{a Q^4}{8 r^6}+\frac{\Lambda r^2}{2}.
\end{equation}
Integrating this equation, the metric function for the electric case is given by
\begin{equation}
f(r)= 1-\frac{2M}{r}+\frac{Q^2}{r^2}-\frac{\Lambda r^2}{3}-\frac{a Q^4}{20 r^6},
\label{gtt}
\end{equation}
where $M$ is the mass of the BH, $Q$ its electric charge, $a$ is the EH parameter
and $\Lambda$ is the cosmological constant that can be positive or negative.
In (\ref{gtt}) the case $a=0$ corresponds to the Reissner-Nordstrom (RN) solution 
with cosmological constant (RN-$\Lambda$).
We recall that RN is the static spherically symmetric solution for the coupled gravitational 
and Maxwell electromagnetic fields, with electromagnetic Lagrangian $\mathcal{L}(F)=-F$.
Then the last term is the extra one compared to the metric function of RN-$\Lambda$ BH, 
and its effect is of reinforcing the gravitational attraction, no matter the sign of the charge
$Q$. Then the system's behavior is more Schwarzschild-like than RN-$\Lambda$. The singularity remains at $r=0$
and is stronger and of opposite sign than in RN-$\Lambda$.
In \cite{Rubiera2019} the electric case in the  NLED $F$ framework is treated;
see also \cite{Kruglov2017}.

For $\Lambda =0$ we recover the electrically charged EEH black hole, characterized by
$(M, Q, a)$, previously known  \cite{Amaro}.  Recently, from the static EEH solution 
has been derived a stationary EH solution \cite{Macias2019}.

\begin{figure}[H] \centering
 \includegraphics[scale=0.6]{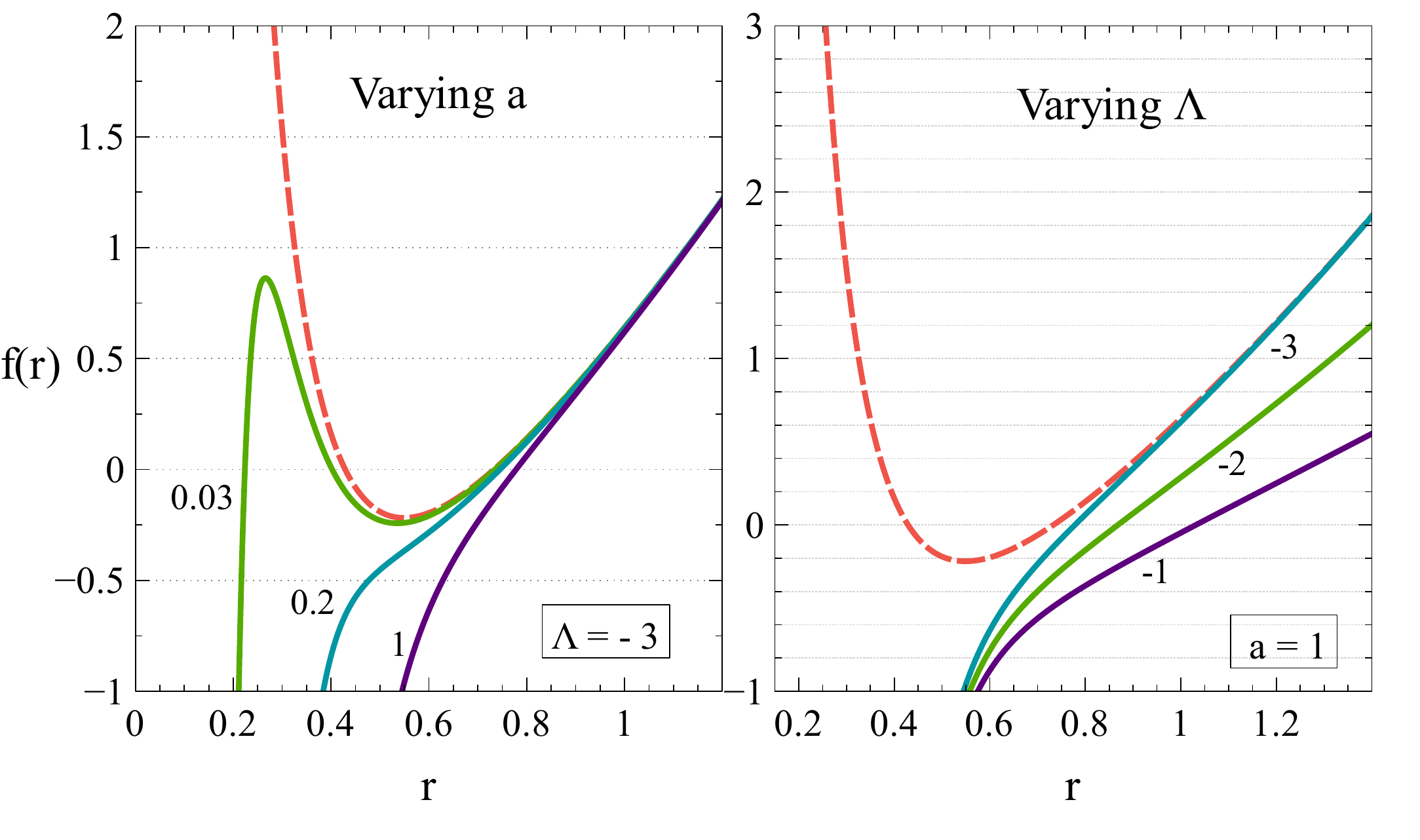}
\caption{\small The metric function $f(r)$ of the EEH-AdS BH  is compared for
fixed values of mass $M=1$ and charge $Q=0.8$. The dashed curves correspond to
Reissner-Nordstrom-AdS BH ($a=0$). The plot to the left is for different values of the EH 
parameter $a$, shown on the curves; while to the right is shown $f(r)$ for different values
of $\Lambda < 0$ (shown on the curves) There are cases with one or three horizons. }\label{fig:frAdS}
\end{figure}

\begin{figure}[H] \centering
\includegraphics[scale=0.5]{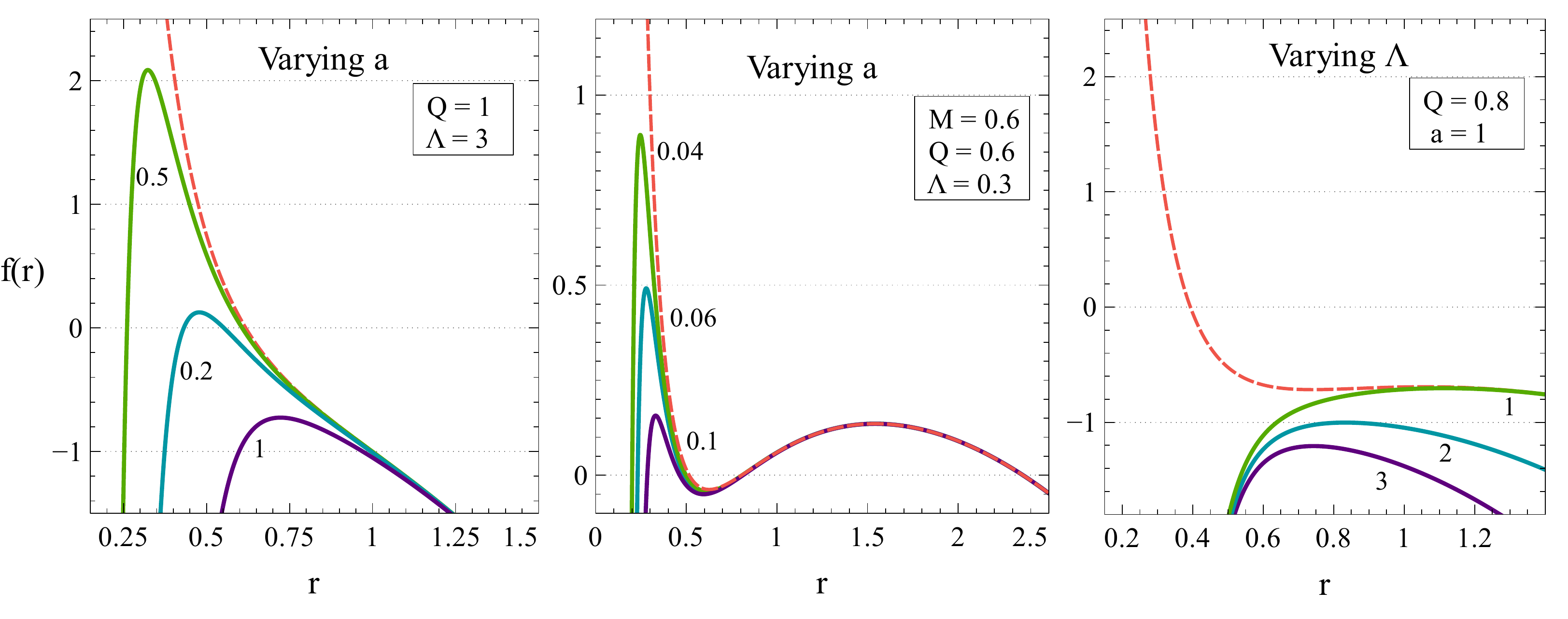}
\caption{\small The metric function $f(r)$ of the EEH-dS BH  is compared
varying $\Lambda >0$ and $a$; the dashed curves are for Reissner-Nordstrom-dS BH ($a=0$).
The plots to 
the left and center are for different values of the EH parameter $a$ as shown on the curves. 
To the right we vary $\Lambda >0 $. As it can be seen, there are cases with none, two, or even 
four horizons (one of them being the cosmological horizon). 
These are for fixed value of mass $M=1$, except for the center figure where M=0.6. }\label{fig:frdS}
\end{figure}

The function $f(r)$ is depicted for different values of the parameters $a$ and
$\Lambda$ in Figs.\ref{fig:frAdS} and \ref{fig:frdS}. In Fig. \ref{fig:frAdS} for a negative $\Lambda$ (anti--de Sitter) while
in Fig. \ref{fig:frdS}  for positive $\Lambda$ (de Sitter).
In general the behavior of $f(r)$ is Schwarzschild-like in contrast  with other
NLED BHs that have a RN behavior but with a screened charge (for
instance Born-Infeld BH \cite{Mann}).
The equation that determines the horizons $f(r_+)=0$ is an eight degree polynomial,

\begin{equation}
\frac{\Lambda}{3} r_+^8 - r_+^6 - Q^2 r_+^4 + 2M r_+^5+ \frac{Q^4 a}{20} =0,    
\end{equation}
then the horizons must be determined  numerically.

A peculiar feature of the solution is the opposite sign of the $Q$ terms in Eq. ($\ref{gtt}$); the minus sign makes the divergences of the solution opposite to those of the RN, that is, the nonlinear term directly affects the sign of the divergence of $f(r)$ and as well the corresponding for the mass and temperature, as we shall see in Sec. $IV$.
However it is important to note that the term $\frac{Q^2}{r^2}$ is always bigger than  $\frac{a Q^4}{r^6}$ for the allowed values of $r$, $r\geq r_+$, and $a$, 0$\leq a \leq \frac{32}{7}Q^2$, [see Eq. (\ref{range_a})], therefore, the dominant behavior of the solution is given by the linear theory. 
\subsection{Electromagnetic  field}

\begin{figure}[H]\centering
    \includegraphics[scale=0.4]{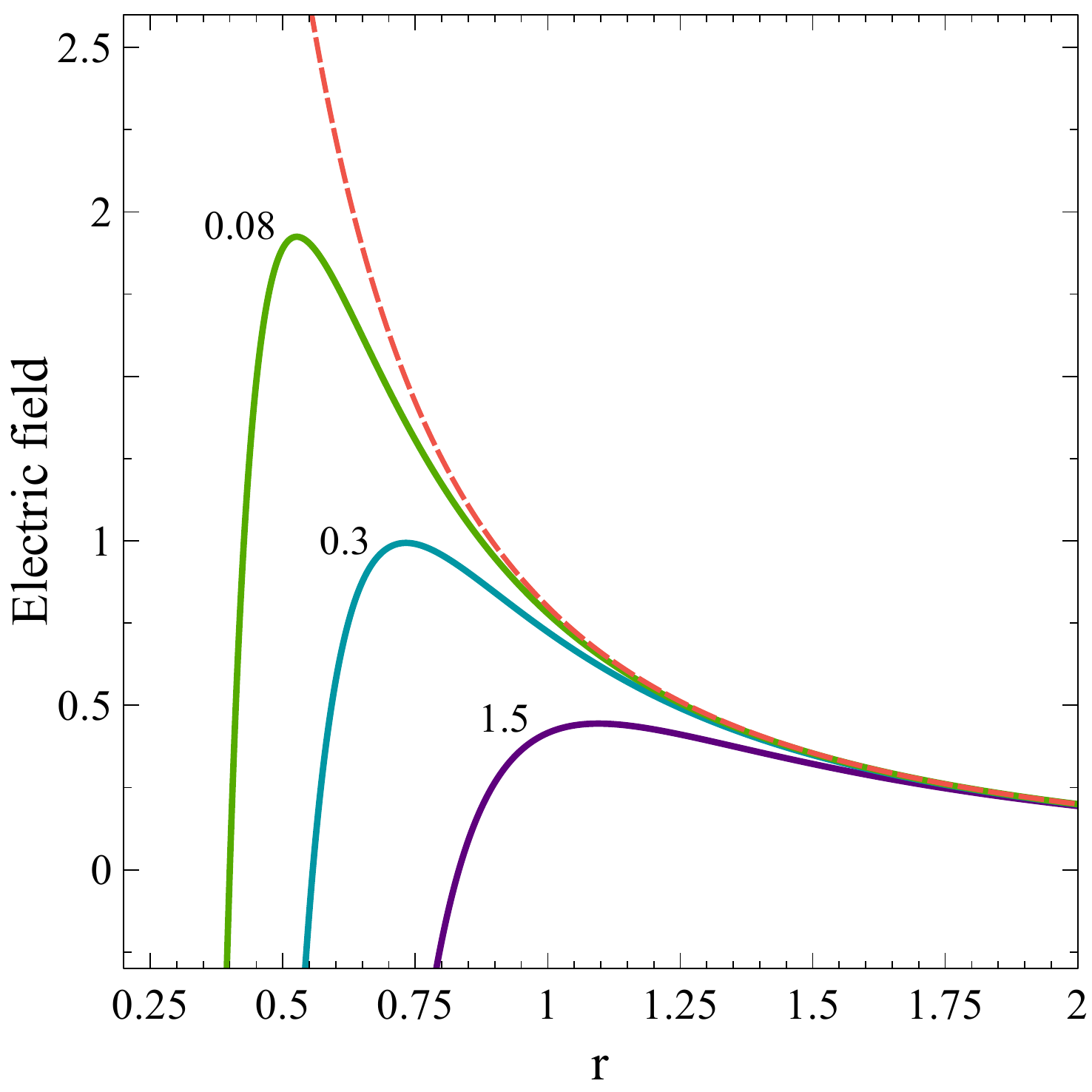}
    \caption{\small The electric field for the EEH-$\Lambda$ BH
for different values of $a$, shown on the curves. Notice that the divergence is
opposite to the Maxwell case RN (dashed curve). In this plot $Q=0.8$.} \label{fig:EF}
\end{figure}

The EH electromagnetic potential $\Phi(r)$ corresponding to the EEH-$\Lambda$ BH is [see Eqs.
(\ref{HamiltonianEH}), (\ref{Pmunu})]
\begin{equation}
\Phi(r)= \int_r^\infty dr P_{tr}\mathcal{H} _P(P, 0)  =
\int_r^\infty dr \frac{Q}{r^2}\left(1-\frac{a\, Q^2}{2r^4}\right)
= \frac{Q}{r} \left(1 -\frac{a\,Q^2}{10 r^4} \right),\label{ElPot}
\end{equation}
and the electric field takes the form
\begin{equation}
    \mathcal{E}(r)= \frac{Q}{r^2} \left(1 -\frac{a \,Q^2}{2r^4} \right),\label{EF}
\end{equation}
where the charge screening is apparent but the divergence at $r=0$ is stronger and
with opposite sign than the corresponding to the RN-$\Lambda$ BH. The field behavior is shown
in Fig. (\ref{fig:EF}) for different values of $a$. The Maxwell
behavior is recovered for $a=0$. For certain radios there are spheres with null electric field
$\mathcal{E}(r_o)=0$.
\subsection{Effective potential}

We present a brief analysis of the effective potential for test particles in the
EEH-$\Lambda$ BH  just to
remark the differences with the case $\Lambda=0$; a more complete study on
geodesics of the EEH BH with $\Lambda=0$ can be found in \cite{Amaro}.
In a static and axisymmetric spacetime, the energy $E$ and the angular momentum
$L_z$ of a test particle of mass $m$ and momentum $P^i$ are
conserved quantities, with $P^i =m\frac{dx^i}{d\lambda}$,
$x^i=t, r,\phi$ and $\lambda$ being the affine parameter along the geodesics,
\begin{eqnarray}\label{conserved}
 E\,&=&-P_t = f(r)m\dot t, \nonumber\\
L_z&=&P_\phi= r^2 \sin^2\theta m \dot\phi,
\end{eqnarray}
where the function $f(r)$ is defined in Eq.
(\ref{gtt}) and the dot represents the derivative with respect to $\lambda$.\\

\begin{figure}[H]\centering
\includegraphics[scale=0.4]{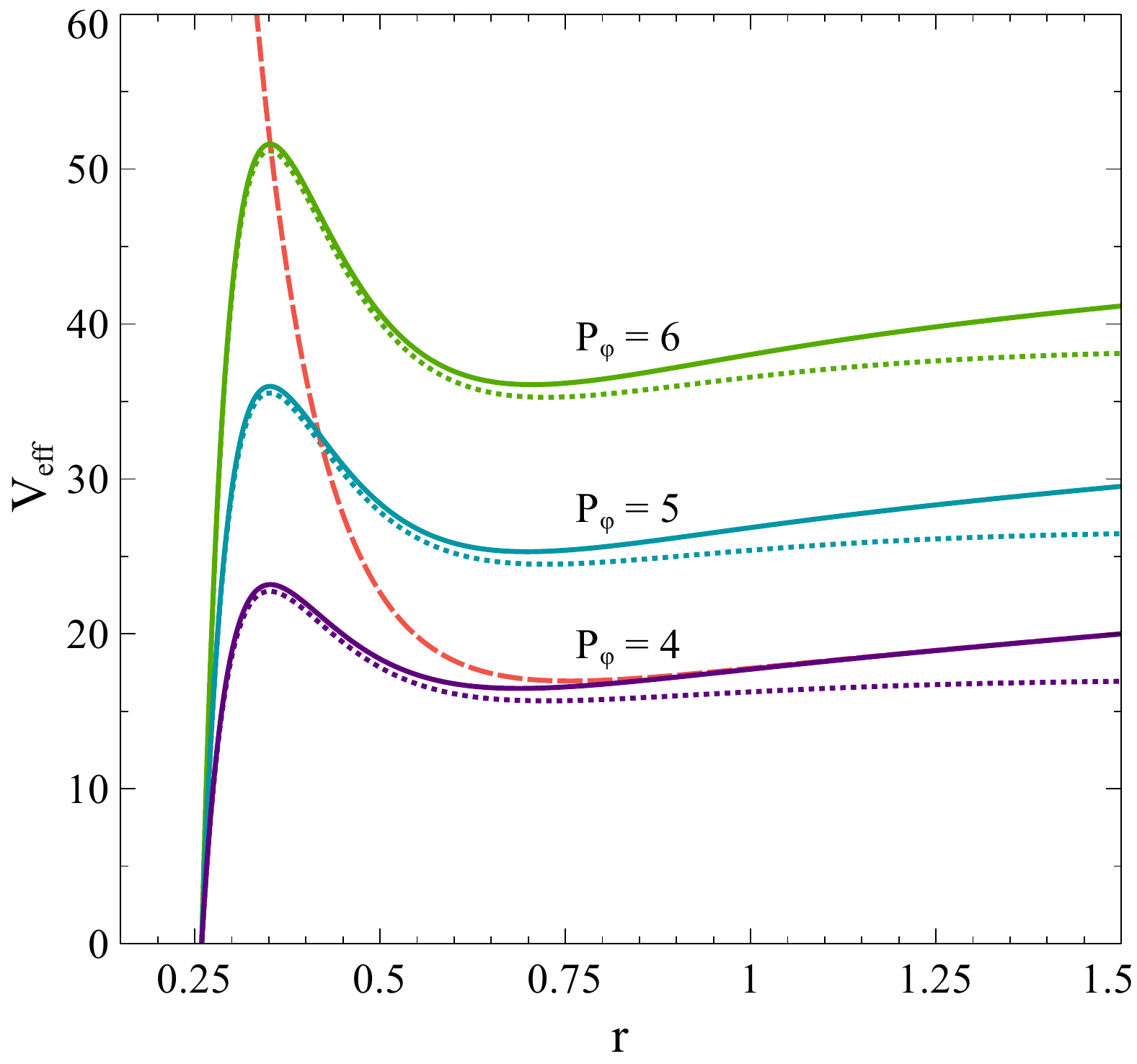}
\caption{\small The effective potential for different values of the angular momentum of the test particle, $P_\phi$, in the vicinity of the EEH-AdS BH. The dotted curves correspond to massless particles ($\omega=0$), while the continuous ones to massive
particles ($\omega= -1$). $ V_{\rm eff}=0$ at the
horizon of the BH, $r_+=0.25$ in this plot.
The dashed curve corresponds to the RN-AdS effective potential.
The  fixed parameters are  $M=1,\; Q=1,\; a=0.3$ and $\Lambda=-3$.}\label{Veff2}
\end{figure}

The geodesic equations  for $t(\lambda)$ and $\phi(\lambda)$ are 
\begin{eqnarray}
\dot t &=&\;\frac{E}{f(r)},\nonumber\\
\dot \phi &=& \;\frac{P_\phi}{r^2 \sin^2\theta},\label{dots}
\label{dot_r}
\end{eqnarray}
where from now on $E$ and $P_\phi$ are the energy and angular momentum per unit of
mass of the test particle. 
Restricting the movement to the equatorial plane $\theta= \pi /2$, and using 
Eqs. (\ref{dots}) we obtain the radial equation $r(\lambda)$,
\begin{equation}
 {\dot r}^2 =E^2-f(r)\left(\frac{P_\phi^2}{r^2}-\omega\right).
\end{equation}
The constant $\omega$ is determined from the scalar product
$g^{ij}\frac{dx^i}{d\tau} \frac{dx^j}{d\tau}=\omega$, and can take values
$\{$-1, 0, 1 $\}$, corresponding to time, null and space-like geodesics respectively.
Comparing with ${\dot r}^2= {E}^2-V_{\rm eff}^2$, we can identify the effective
potential term as

\begin{equation}
 V_{\rm eff}^2=f(r)\left(\frac{P_\phi^2}{r^2}-\omega\right), \label{Veff}
\end{equation}
that is depicted in Fig. \ref{Veff2} for different values of the angular momentum of the test particle.
Note that for both massive and massless particles,  stable circular orbits (minimum
of $ V_{\rm eff}$) as well as unstable circular orbits (maximum of $ V_{\rm eff}$)
are possible. In the absence of a cosmological constant there are not stable circular orbits for massless test particles i.e.
$V_{\rm eff}$ does not present a minimum,  
this being the main difference respect to the EEH BH with $\Lambda=0$. 

\section{Thermodynamics of the EEH-AdS BH}

This section is devoted to the thermodynamics of the EEH-AdS BH.
From now on we shall consider $\Lambda$
being negative, i.e. the space is anti--de Sitter (AdS), and consider $\Lambda$ as the
thermodynamic pressure. We shall explore the
similarities and differences with its linear version, the RN-AdS BH.
We also refer to the study of the NLED BI BH in \cite{Mann}, but bear in mind 
that in the BI case the linear limit RN-AdS is reached when the BI parameter goes to infinity,
while in the EH theory that limit corresponds to the vanishing of the EH parameter $a=0$.
Therefore at first sight when comparing the behaviors of the BI and EH BHs they appear different, 
but qualitatively are very alike.
We also show  the
concordance between the first law and the Smarr formula for the BH mass. Moreover,
we determine the critical points, phase transitions and up to what point the
critical exponents of the EEH-AdS BH reproduce the Van der Waals fluid ones. 

For the EEH-AdS BH, metric (\ref{metric}) with Eq. (\ref{gtt}), 
the temperature $T$, pressure $P$, volume $V$ and entropy $S$ are
given by

\begin{equation}
T=\frac{f'(r_+)}{4\pi}=\frac{1}{4\pi r_+}\left(1-\frac{Q^2}{r_+^2}+\frac{a
Q^4}{4r_+^6}-\Lambda r_+^2\right),\label{temp}
\end{equation}
\begin{eqnarray}
P=-\frac{\Lambda}{8\pi}, \hspace{1.2cm} V=\frac{4}{3}\pi r_+^3, \hspace{1.2cm} S=
\pi r_{+}^2, \label{PVS}
\end{eqnarray}
where $r_{+}(M, Q, a, \Lambda)$ is the outer horizon.

\subsection{First law and Smarr's formula for the EEH-AdS-BH}

The possibility of formulating the corresponding Smarr relation \cite{Smarr73} in nonlinear 
electrodynamics has been addressed previously \cite{Breton2005}, \cite{Smolic2018}.
In \cite{Smolic2018} a generalized Smarr formula is
derived, via geometric arguments, valid
for any stationary axially symmetric black hole with the electromagnetic field defined
by some of the nonlinear models. However, it is not considered the EH BH in the extended
space. 

In order to obtain the Smarr formula for the EEH-AdS BH, we perform a scaling analysis
through the Euler's theorem, which states that if a function $f(x,y)$ obeys
$f(\alpha^p x,\alpha^q x)= \alpha^r f(x,y)$, then it follows that
\begin{equation}
    r f(x,y)= p x \left(\frac{\partial f}{\partial x}\right) +q y
\left(\frac{\partial f}{\partial y}\right);\nonumber
\end{equation}
this is a two variable formulation of the theorem, but of course is valid
in a higher dimensional space.

The mass $M$ of the EEH-AdS BH must be considered as a function of the entropy $S$, the
pressure $P$, the charge $Q$ and the EH parameter $a$. Let us
see the units of the thermodynamic quantities, $[M]=L$,  $[S]=L^2$, $[P]=L^{-2}$,  $[Q]=L$, 
$[a]= 1 / [\mathcal{E}] = L^2$, where $L$ represents  units of length. Then,
according to the previous theorem, the mass satisfies  
\begin{equation}
    M(S,P,Q,a) = 2S\left( \frac{\partial M}{\partial S}\right) -2P
\left(\frac{\partial M}{\partial P}\right)+Q \left(\frac{\partial M}{\partial
Q}\right)+2a \left(\frac{\partial M}{\partial a}\right) .\label{M}
\end{equation}
The partial derivatives of $M$ can be found by considering the first law of AdS
BHs, related to energy conservation,  that includes variations in the cosmological constant \cite{Kastor} 
\begin{equation}
    dM= TdS+VdP+\Phi dQ,\label{firstlaw}
\end{equation}
and using the fact that $M$ is a perfect differential, thus obtaining
\begin{equation}
    \left(\frac{\partial M}{\partial S}\right)=T, \hspace{.5cm} \left(\frac{\partial
M}{\partial P}\right)=V, \hspace{.5cm}
    \left(\frac{\partial M}{\partial Q}\right)=\Phi. \hspace{.5cm}\nonumber
\end{equation}

Therefore, the Smarr formula that we propose takes the form
\begin{equation}
    M= 2(TS-VP+\mathcal{A} a)+\Phi\, Q,\label{Smarr}
\end{equation}

where $\mathcal{A}\equiv \left(\frac{\partial M}{\partial a}\right)$ is the
conjugate of the EH parameter $a$. We determine  $\mathcal{A}$ directly by 
substituting Eqs. (\ref{ElPot}), (\ref{temp}) and (\ref{PVS}) into the Smarr
formula and also using the fact that $f(r_+)=0$,
\begin{equation}
    \mathcal{A}= -\frac{Q^4}{40 r_+^5}.\label{A}
\end{equation}

In such a way the EEH-AdS BH Smarr formula  is given by
\begin{equation}
M= \frac{r_{+}}{2}+\frac{Q^2}{2r_{+}}-\frac{a\,Q^4}{40 r_{+}^5}-\frac{\Lambda
r_{+}^3}{6},\label{mass1}
\end{equation}
considering that $r_+$ is a function of the  EEH-AdS BH parameters, $r_+(M,Q,a, \Lambda)$.

Let us check now the consistency of the first law and the corresponding Smarr
formula (\ref{Smarr}).
Choosing a convenient path of integration such that the energy of the EEH black hole
is defined by its components \cite{Smarr73}: the surface energy $E_s$ at the horizon and
the electromagnetic energy $E_{em}$, we have

\begin{eqnarray}
          E_s&=&\int_0^r T(r',Q=0) 2\pi r'dr',\nonumber\\
          E_{em}&=&\int_0^Q \Phi(r,Q',a) dQ', \hspace{.8cm} r, \;a\mbox{
fixed},\nonumber
\end{eqnarray}
where $T$ is the trace of the energy momentum tensor in Eq. (\ref{emtensor})
Therefore, according to the {\it first law}, the mass is determined by 
$M=E_s + E_{em}$, i.e.
\begin{equation}
M= \frac{r_{+}}{2}+\frac{Q^2}{2r_{+}}-\frac{a\,Q^4}{40 r_{+}^5}-\frac{\Lambda
r_{+}^3}{6},
\end{equation}

which is in agreement with the {\it Smarr formula} as proposed in Eq. (\ref{Smarr}), and the
RN-AdS mass is recovered if $a=0$. \\ 

Let us point out to a possible interpretation of the EH parameter $a$ and its conjugate $\mathcal{A}$. Considering that a change $da$ in $a$ will change the electrostatic  energy of the system in $\mathcal{A} da$, this suggests an interpretation of the term $ a \mathcal{A}$ in the Smarr formula as the electrostatic energy that arises from the vacuum polarization and is stored in the horizon of the black hole. This is also an analogy to a polarizable medium
in which the energy is given by the product of the polarization times the electric field. The dimensional analysis is in agreement with this interpretation:
$a$ has dimensions of polarization, $[a]= L^2$ while  $\mathcal{A}$ has dimensions of electric field $[\mathcal{A} ]=L^{-1}$, that together gives the right dimensions for the enthalpy, $[M]=L$ with units of energy.

\subsection{Thermodynamical stability}

The thermodynamical (in)stability of a black hole can be determined by the behavior of the specific heat $C_Q$. In this subsection we show that the effect of introducing the EH parameter $a$ is the appearance of a phase transition.

In \cite{Breton2015} it was shown that, for any Lagrangian describing nonlinear electromagnetism, a divergence with a change of sign in the heat capacity $C_Q$ at constant charge corresponds to a change from stability to instability (or vice versa) that indicates a phase transition.

The heat capacity at constant charge is given by
\begin{equation}
     C_Q=-2\pi r_+^2 \frac{4r_+^6-4Q^2r_+^4+aQ^4-4\Lambda r_+^8}{4r_+^6-12Q^2r_+^4+7aQ^4+4\Lambda r_+^8}.
\end{equation}

\begin{figure}[H]
    \centering
    \includegraphics[scale=0.5]{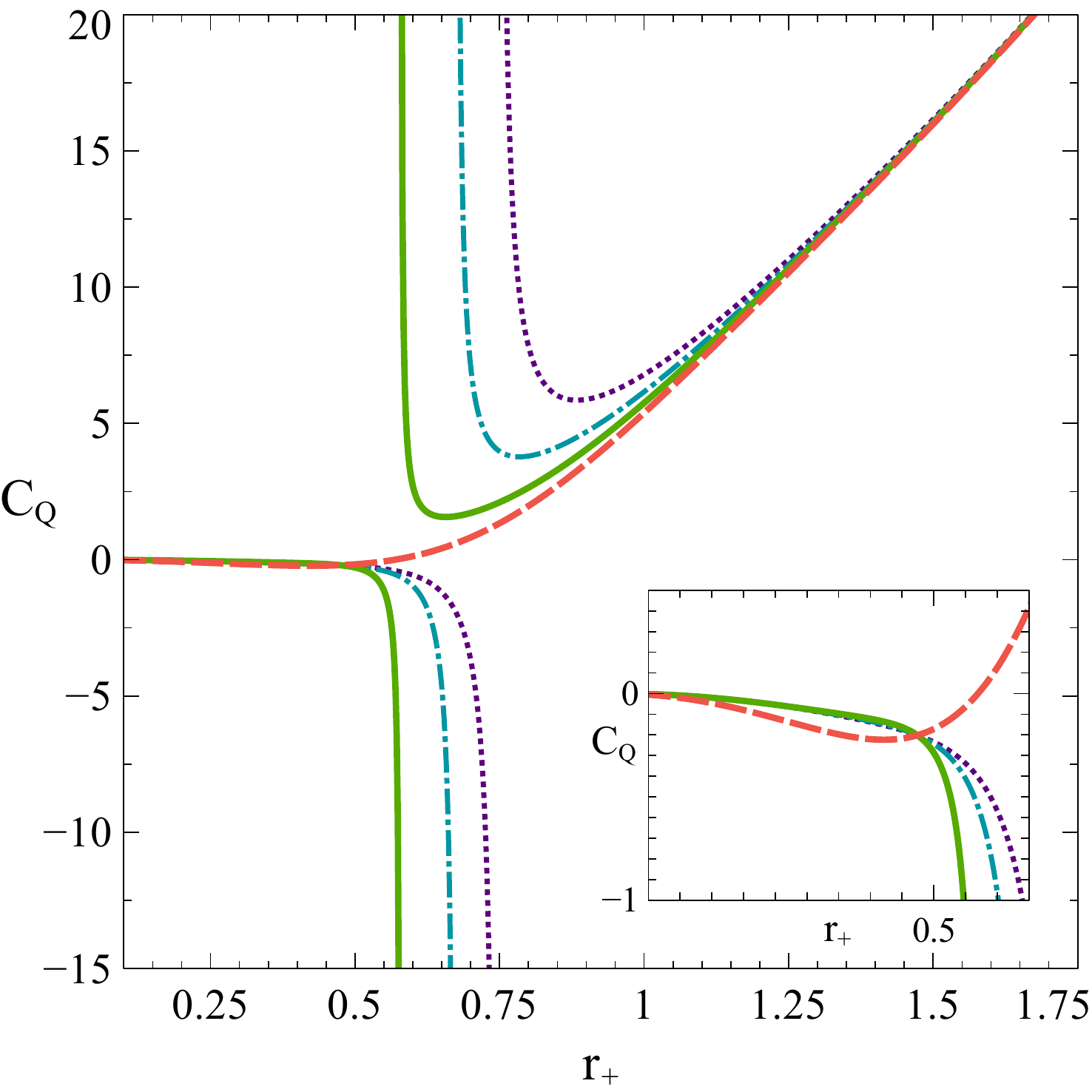}
    \caption{\small The heat capacity at constant $Q$, $C_Q$, is shown as a function of the horizon radius $r_+$, for several values of the EH parameter $a$: the solid (green) curve is for $a=0.3$, the dash-dotted (blue) is for $a=0.6$ and the dotted (purple) is for $a=1$. The dashed curve is for RN-AdS $a=0$. A negative $C_Q$ corresponds to unstable configurations while a positive $C_Q$ is interpreted as stable.
    As $a$ increases the phase transition unstable-stable occurs at greater values of the horizon radius $r_+$.
    The fixed parameters are $Q=0.8$, $\Lambda=-3$. In the small box, a zoom of the region near $r_+=0$ is shown. }
    \label{fig:CQ}
\end{figure}

It is illustrated in Fig. \ref{fig:CQ} and compared with the corresponding to RN-AdS BH. The discontinuity in $C_Q$ and a change of sign are observed, indicating instability of the black hole in the canonical ensemble.

A negative specific heat $C_Q < 0$ corresponds to unstable configurations while a possitive $C_Q$ is interpreted as stable ones. Therefore the introduction of the EH parameter induces a change from instability to stability, through a phase transition. The change in RN-AdS occurs as well but in a different way, as illustrated by the behavior of $C_Q.$

In the forthcoming subsections we shall see that this instability leads to a phase transition. 

\subsection{Critical behavior of the EEH-AdS BH thermodynamics.}

The previous analysis allows us to  write down a EEH-AdS BH {\it equation of state}
through Eq. (\ref{temp}) as

\begin{equation}
P= \frac{T}{2r_+}-\frac{1}{8\pi r_+^2}+\frac{Q^2}{8\pi r_+^4}-\frac{a Q^4}{32\pi
r_+^8}.
\end{equation}

This equation can be written also in terms of the specific volume $v=2 L^2 r_+$,
taking $L=1$, the equation of state reads
\begin{equation}
  P= \frac{T}{v}-\frac{1}{2\pi v^2}+\frac{2 Q^2}{\pi v^4}-\frac{8 a Q^4}{\pi v^8}.
\label{p(v)} 
\end{equation} 
As it is expected, the equation of state for RN-AdS BH is
recovered for $a=0$, (see {\it e.g.} \cite{JoyDas}).

The location of the phase transition is determined by the critical points that can
be seen
from the $P-v$ diagram and are determined demanding the conditions  $\partial P/
\partial v=0$ and $\partial^2 P/ \partial v ^2=0$.
These conditions lead to a third degree equation for the critical volume squared, $x\equiv v_c^2$
 \begin{equation}
     x^3-24 Q^2 x^2 +448 a Q^4=0.\label{cubic}
 \end{equation}

\begin{figure}[H] \centering
\includegraphics[scale=0.45]{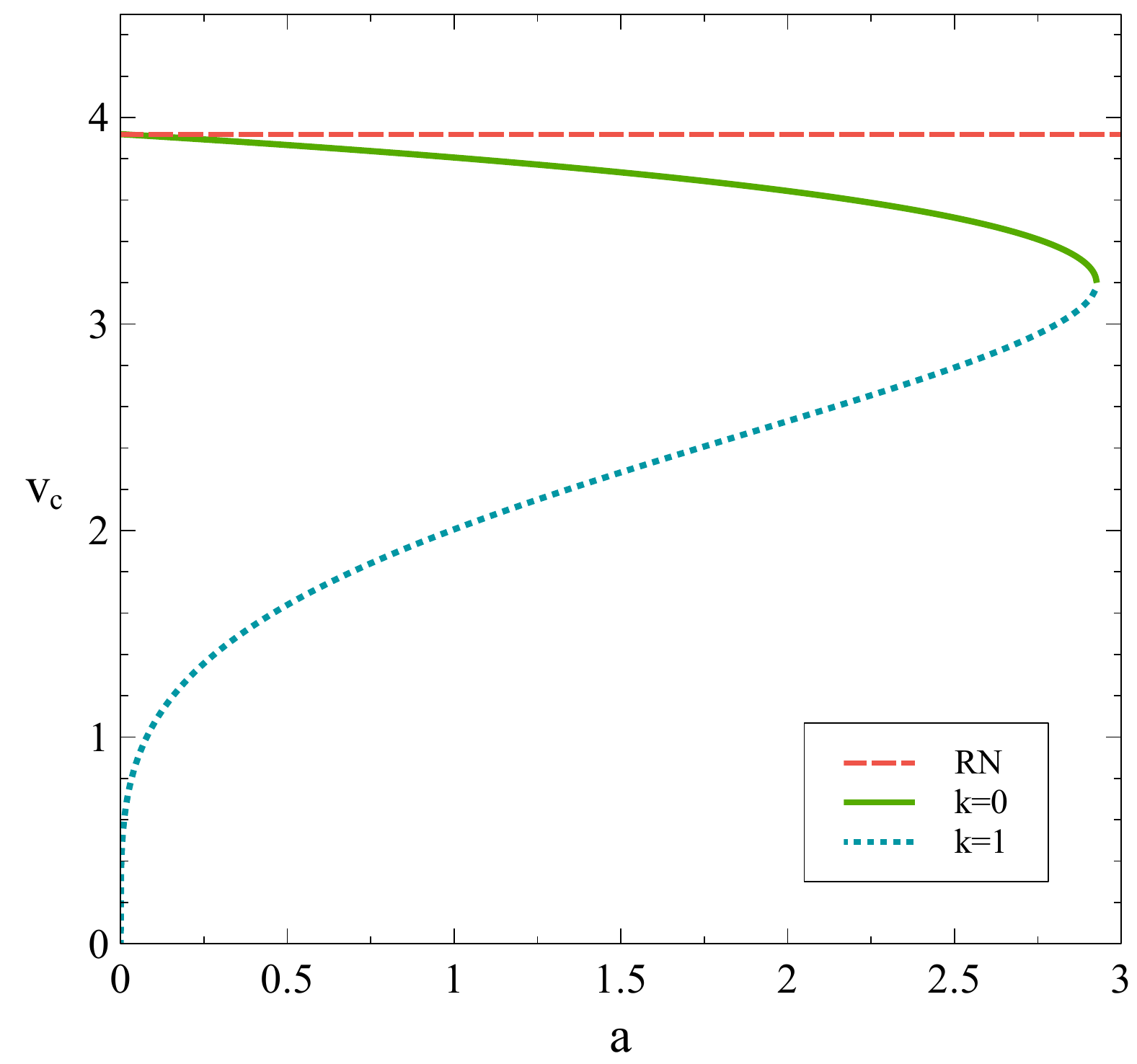}
\caption{\small It is shown the critical volume $v_c$ vs. the EH parameter, $a$, for the EEH-AdS BH. 
There are two branches ($k=0$ and $k=1$); the branch $k=0$ is the
closest to the linear one (RN-AdS BH), shown by the dashed line. The branch $k=1$ (dotted curve) 
differs more from the Maxwell regime, being the fully nonlinear electromagnetic branch. 
The point where the
solid and the dotted curves meet corresponds to the upper bound for the EH parameter
$a=\frac{32}{7} Q^2$. In this plot  $Q=0.8$.}
\label{fig:vc}
\end{figure}

 The cubic Eq. (\ref{cubic}) admits three real roots if
  \begin{equation}
     0\leq a \leq \frac{32}{7} Q^2,\label{range_a}
 \end{equation}
 in which case, the roots are given by
  \begin{equation}
     x_k= 8Q^2 \left( 2\cos{\left[\frac{1}{3}\arccos{\left( 1-\frac{7
a}{16Q^2}\right)-\frac{2\pi k}{3}}\right]}+1\right), \hspace{.7cm} k=0,1,2. 
\label{xk}
\end{equation} 
However, it is also required that $x_k$ be positive, since $v_c$ is a volume, ($v_c=\sqrt{x_k}$), that is never satisfied for $x_2$, therefore, only $x_0$ and $x_1$ are meaningful to our system. The valid roots $x_0$ and $x_1$ are shown in Fig. \ref{fig:vc}.

It is worth mentioning that if $a>\frac{32}{7} Q^2$, the three solutions of Eq. (\ref{cubic}) are nonphysical as a volume (one is real but negative, the other two being complex conjugate).
 From these solutions for $v_c$, the critical temperature $T_c$ and pressure $P_c$ are given by
 \begin{eqnarray}
     T_{ck} &=& \frac{1}{\pi v_c} \left[ 1-\frac{8Q^2}{x_k}+\frac{64 a
Q^4}{x_k^3}\right],\nonumber\\
     P_{ck} &=& \frac{1}{\pi x_k} \left[ \frac{1}{2}-\frac{6Q^2}{x_k}+\frac{56 a
Q^4}{x_k^3}\right], \hspace{.8cm} k=0,1.\label{Tc,Pc}
 \end{eqnarray}
 
\begin{figure}[H]
    \centering
    \subfigure{\includegraphics[scale=0.5]{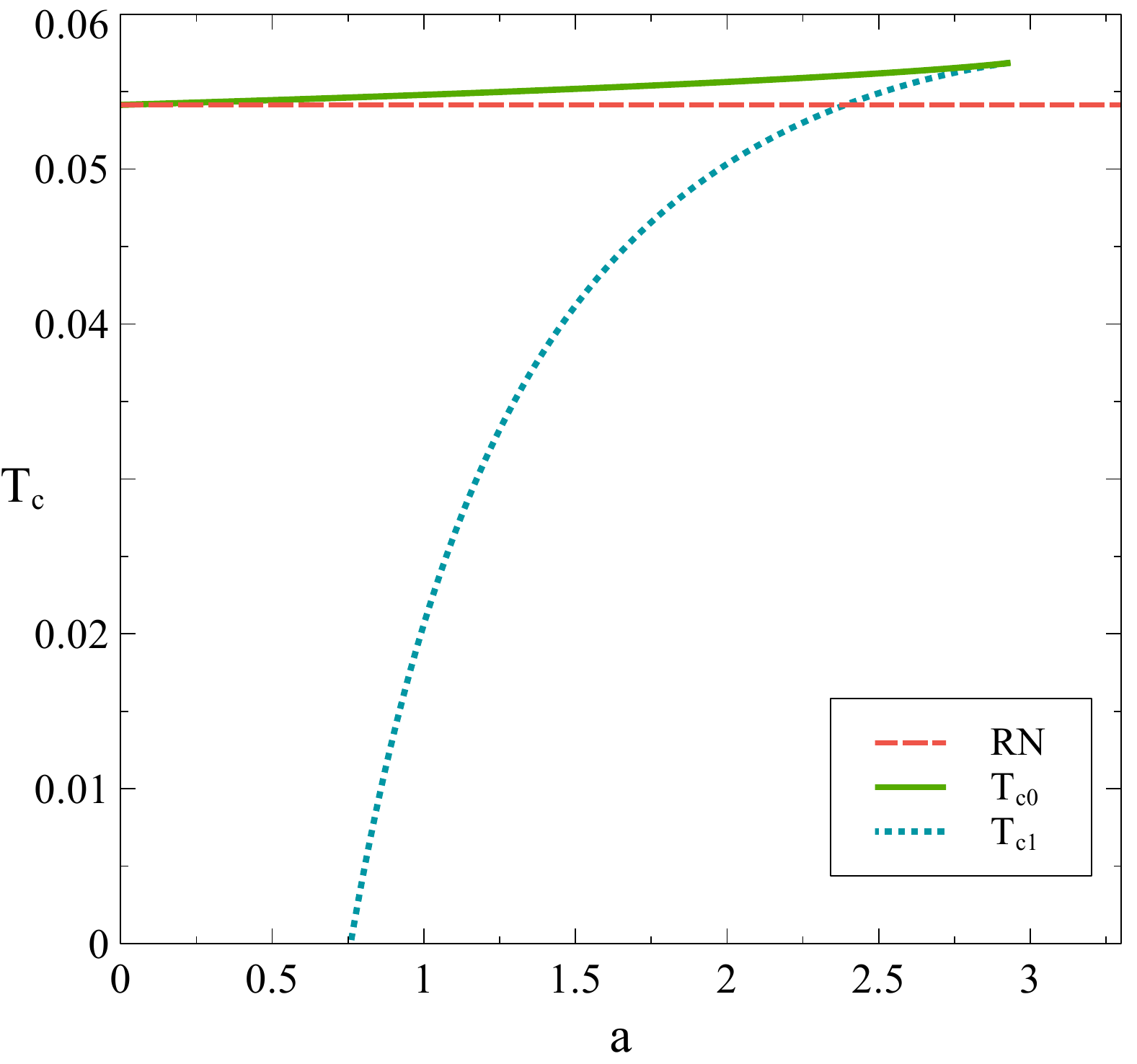}}
    \subfigure{\includegraphics[scale=0.5]{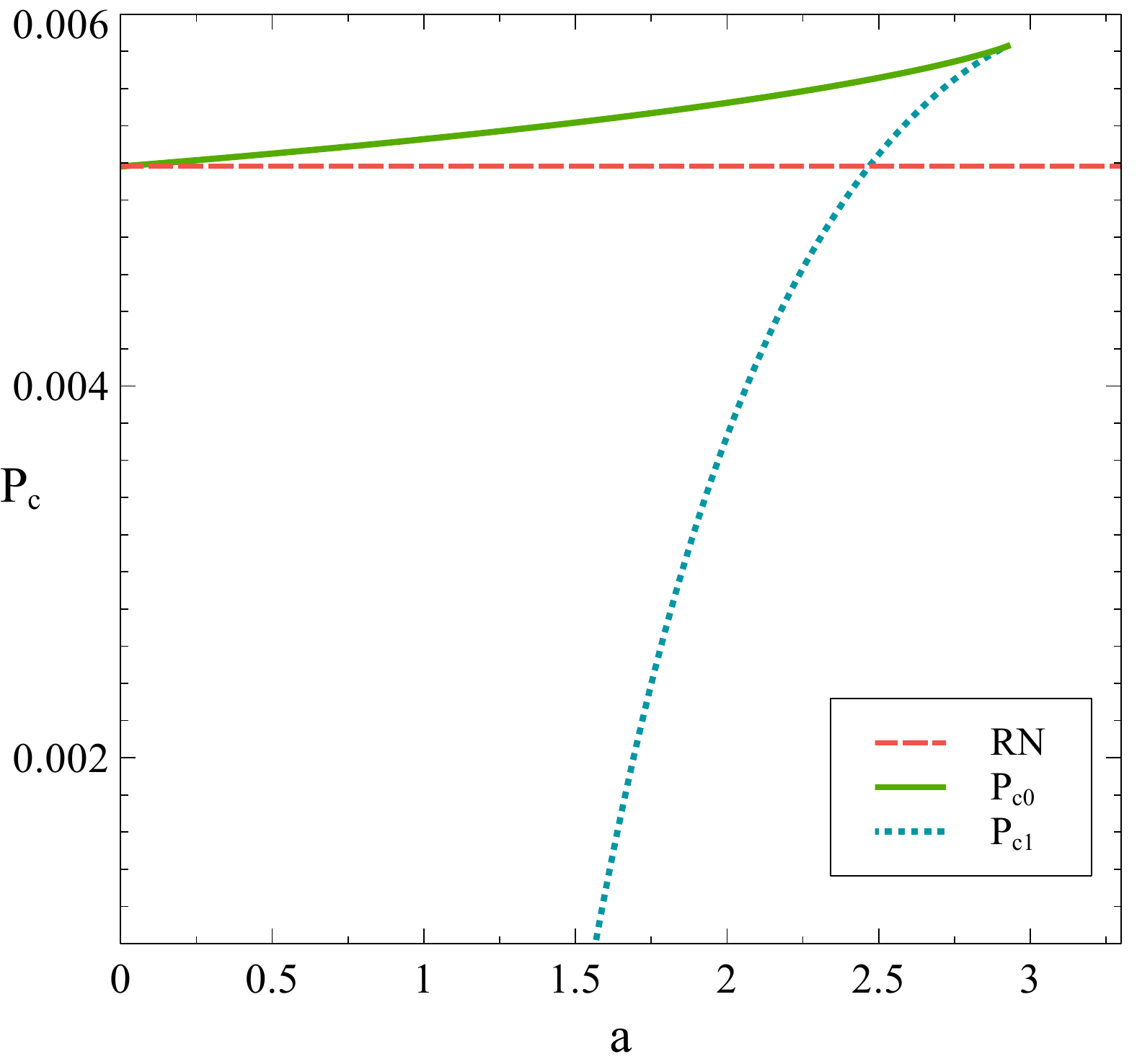}}
    \caption{\small It is shown the critical temperature (left) and pressure (right) as a function of the EH
parameter $a$ for the EEH-AdS BH; in both cases the branch $k=0$ (solid curve) is the closest to the linear one (RN-AdS BH), shown by the dashed
line. The branch $k=1$ (dotted curve) differs more from the Maxwell regime, being the fully nonlinear electromagnetic
branch. The cusp corresponds to the turning point in the critical volume
graphic ($x_0=x_1$) where $a=32Q^2/7$. In this plot $Q=0.8$.}
\label{fig:Tc}
\end{figure}

The two branches of the critical temperature $T_{c0}$ and $T_{c1}$ as well as the critical pressure $P_{c0}$ and $P_{c1}$ are depicted in
Fig. \ref{fig:Tc}; in both cases the critical quantities increase with the EH parameter $a$. Notice that  there exists a restriction over $a$ in the branch $P_{c1}$ in order to have a positive pressure; however, the phase transition occurs in reference to the branch $k=0$, as we can see below. For this reason we will not specify details about the branch $k=1$.

\begin{figure}[H] \centering
\includegraphics[scale=0.5]{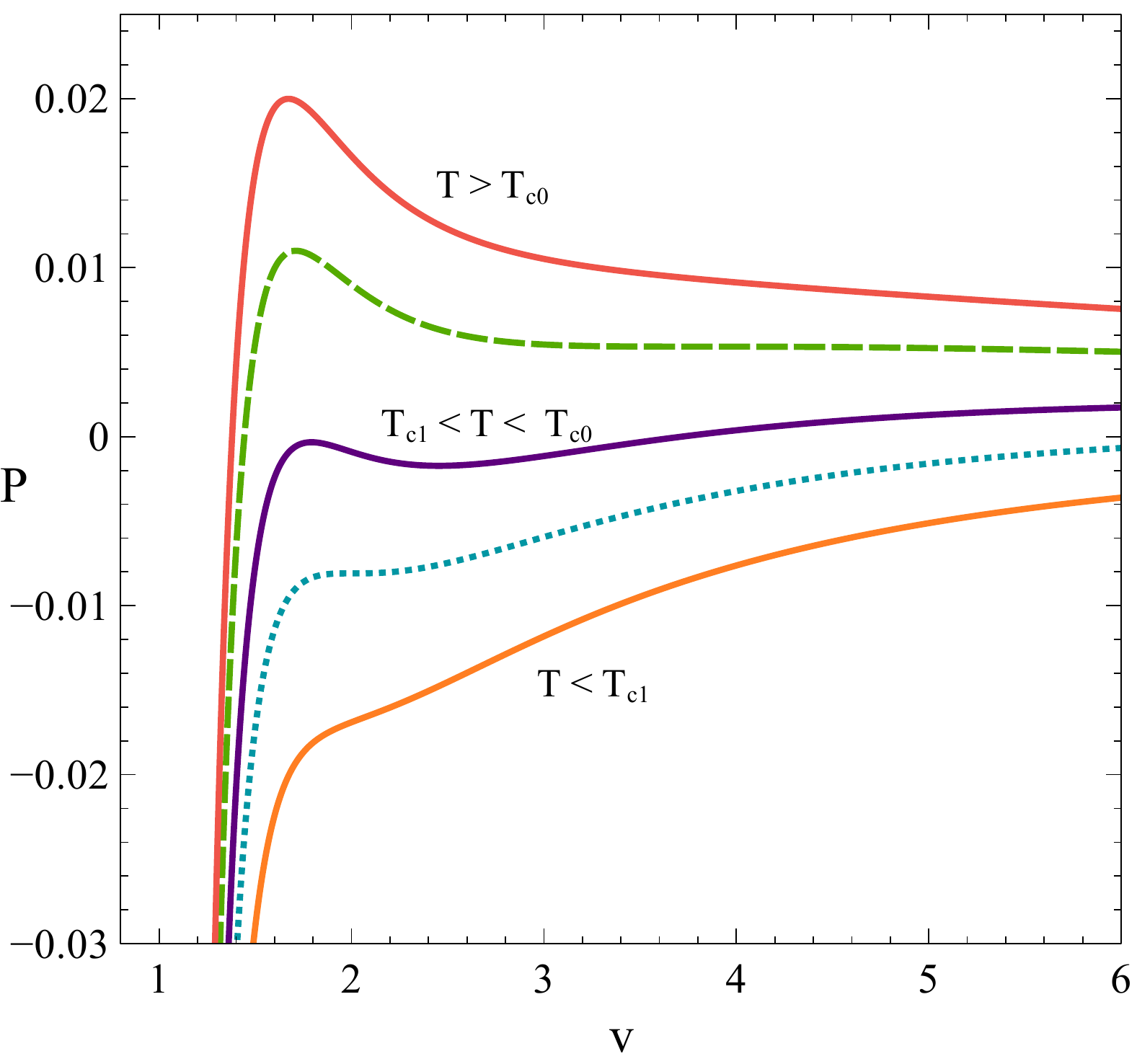}
\caption{\small $P$-$v$ diagrams for fixed temperatures for the EEH-AdS BH.. 
The dashed curve is for $T=T_{c0}$ ($k=0$ branch) and the dotted one is for
$T=T_{c1}$ ($k=1$ branch). The fixed values are $Q=0.8$ and $a=1$. }
\label{fig:PV}
\end{figure}

In Fig. \ref{fig:PV} are shown 
the $P$-$v$ diagrams, taking as a reference the critical values of temperature, corresponding to the 
two roots $x_0$ and $x_1$ of the critical volume. 
For the isotherm corresponding to $T= T_{c0}$, there exist maximum and minimum 
that indicates the possible existence of two  phase transitions around that temperature.
For the isotherms $T\leq T_{c1}$ the pressure is negative in the region where the phase transition could occur. Even though negative pressure could be associated to the vacuum polarization, we left
this analysis for a forthcoming work. Therefore in what follows we 
focus on the branch $k=0$.
We can take small values of the parameter $a$ in order to approximate  the critical
points in Eq. (\ref{Tc,Pc}). Taking into account that in this limit the value of $x_1$ leads to a zero
volume  we only consider in our analysis the root $k=0$,

\begin{equation}
\lim_{a\to 0} x_0 = 16\, Q^2 \lim_{a\to 0} \cos{\left[\frac{1}{3}\arccos{\left(
1-\frac{7 a}{16Q^2}\right)}\right]} + 8Q^2=24 Q^2;
\end{equation}
thus, in this limit the  critical quantities tend to 
\begin{eqnarray}
v_c&\approx& 2\sqrt{6}\,Q\nonumber \\
T_c&\approx& \frac{1}{3\sqrt{6}\pi Q}+ \frac{a}{432\pi Q^3},\label{Crit}\\ 
 P_c &\approx& \frac{1}{4*24 \pi Q^2}+\frac{7a}{41472 \pi Q^4}. \nonumber
\end{eqnarray}
Therefore  the critical ratio $\rho_c= P_c v_c / T_c$ tends  to
\begin{equation}
    \rho_c \approx \frac{3}{8} -3.02\times 10^{-4}\frac{a}{Q^2}+2.22\times
10^{-4}\frac{a^2}{Q^4}.\label{rho_app}
\end{equation}

To the zeroth order in $a$  is the same critical value in the Van der Waals system,
$\rho_c = \frac{3}{8}$, recovering then the RN-AdS BH value as well. To first and second order in $a$
the difference is to the ten thousandth.
 \subsection{Gibbs free energy and phase transitions for the EEH-AdS BH.} 

 The thermodynamic information of a system is completely encoded by its partition
function, which in the semiclassical approximation is associated with the
Euclidean action $I_e$, and the finite part, because the divergences in the
AdS asymptotic region are canceled out by the counterterm method, \cite{Emparan}. We are considering
a canonical ensemble, where the Gibbs
free energy is  $G=\frac{1}{\beta} I_e$, and $\beta$ is the inverse of the BH 
temperature. For the EEH-AdS BH, the Gibbs free energy $G=M-TS$, reads
 
\begin{equation}
G(T,P,v)= \frac{v}{8} +\frac{3Q^2}{2v}-\frac{14 a Q^4}{5v^5}-\frac{\pi v^3
P}{12}.
\end{equation}

\begin{figure}[H] \centering
\includegraphics[scale=0.6]{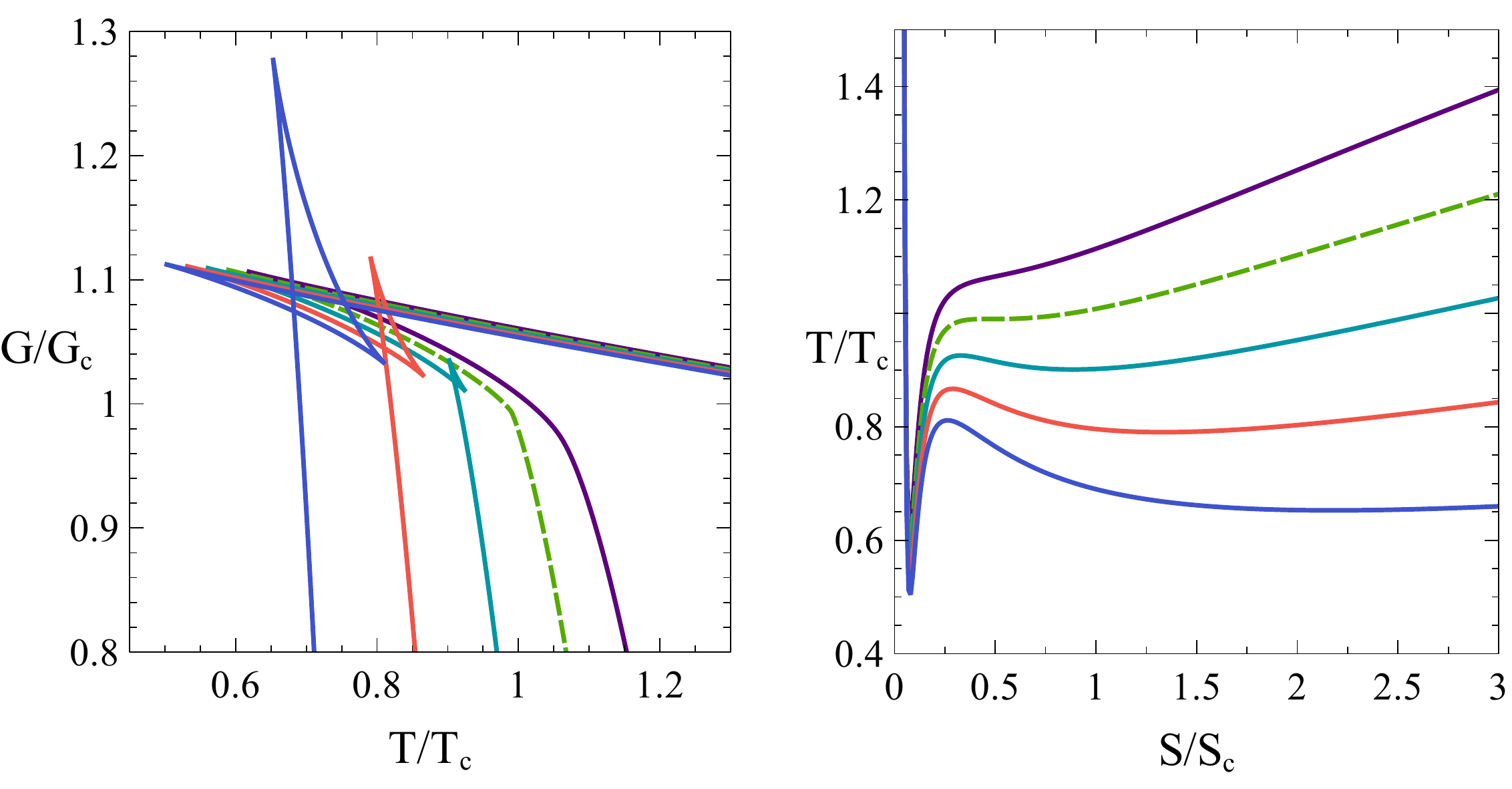}
\caption{\small Isobaric curves in the reduced parameter space of the EEH-AdS BH, for fixed values of
pressure $P/P_c = 0.4,\, 0.6,\, 0.8,\, 1, 1.2$. The dashed curve corresponds to
$P/P_c=1$. Left: Gibss free energy, the pressure
increases from left to right, the first order phase transition occurs for pressures
less than $P_c$. Right: T-S diagram, and the pressure increases from bottom to
top. We have set $Q=1$ and $a=1$.}\label{fig:Gibbs}
\end{figure}

\begin{figure}[H]
\centering
\subfigure{\includegraphics[scale=.37]{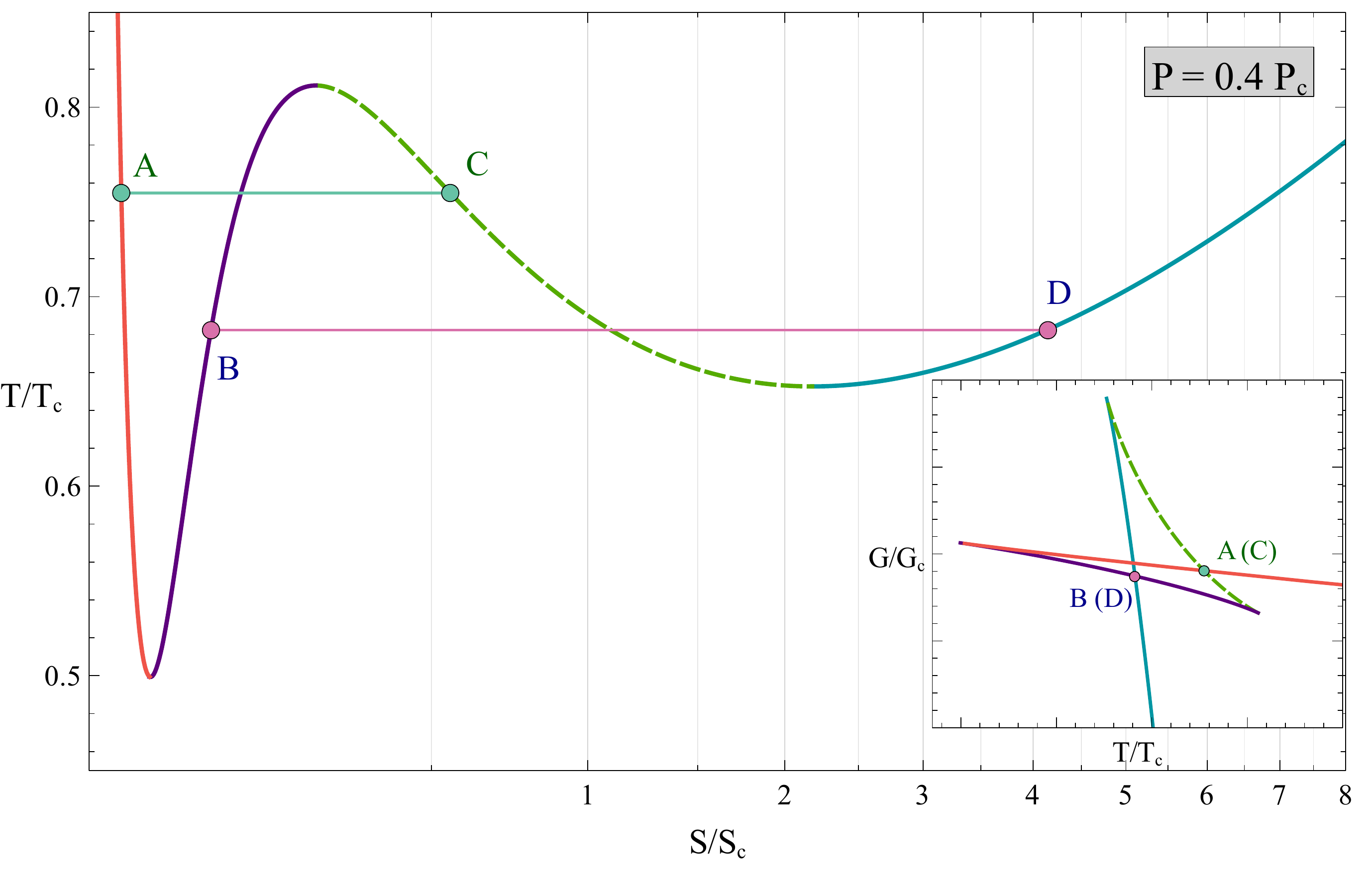}}
\subfigure{\includegraphics[scale=.37]{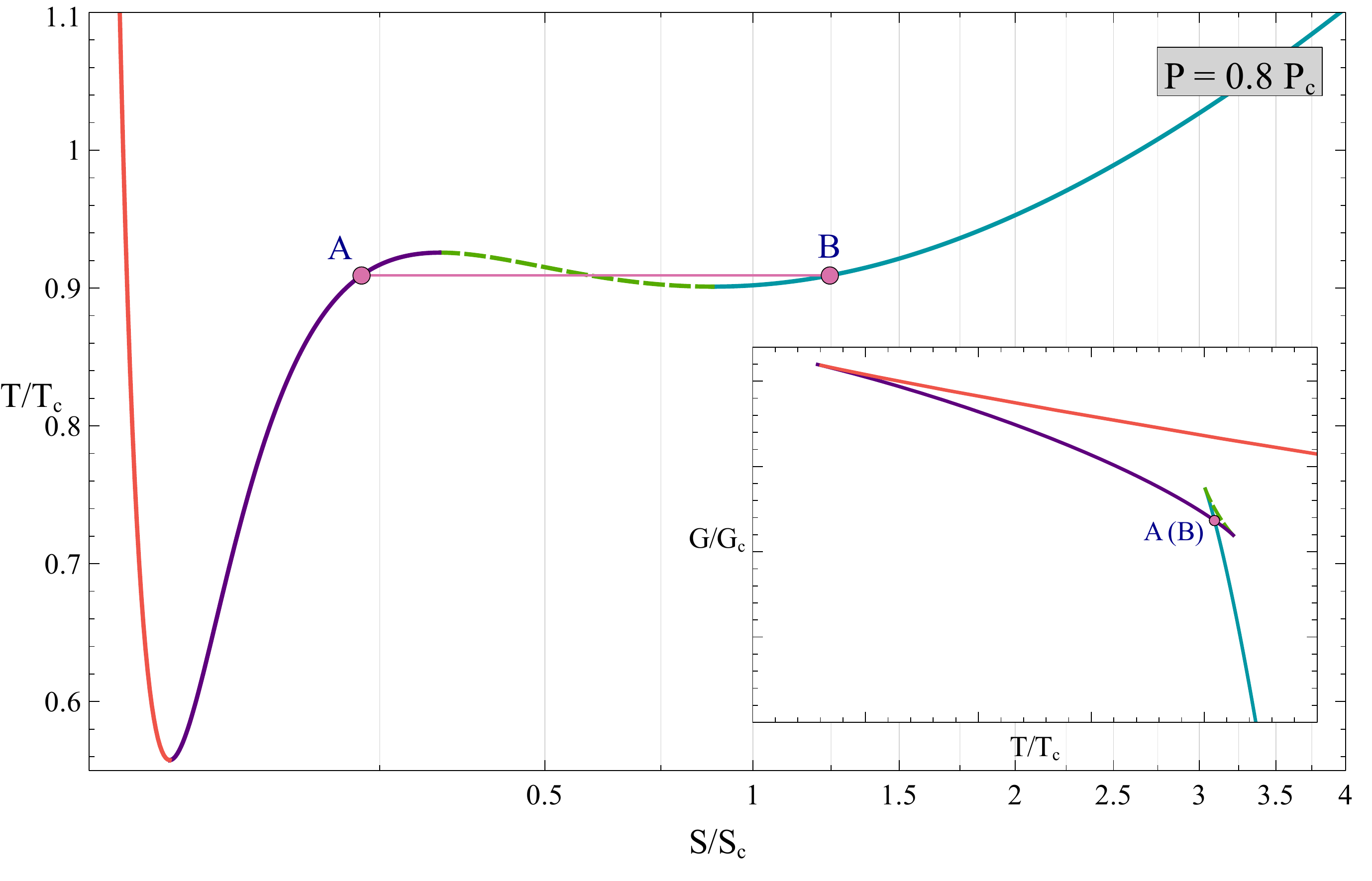}}
\caption{T-S diagrams and Gibbs free energy vs temperature for the EEH-AdS BH; {\it upper:} $P=0.4\,
P_c$, the Maxwell's equal area law is satisfied for the sections $AC$ and $BD$,
where the phase transition occurs. {\it Below:} $P=0.8 \,P_c$ the phase transition occurs
from the point $A$ to $B$, following the Maxwell equal area law.  In the small boxes
are shown the corresponding phase transitions as seen in the Gibbs-Temperature diagrams, 
the dots showing where the transition occurs. In these plots $Q=1$ and $a=1$.} \label{fig:gt4y8}
\end{figure}

In Fig. \ref{fig:Gibbs} we plot the isobars of the the Gibbs free energy and the
$T$-$S$ diagrams for different values of the pressure. In both graphics are shown the reduced
values at the critical point, i.e. $G/G_c$ and $T/T_c$, etc. with 
$G_c = G(T_c,P_c,v_c)$ and $S_c=S(v_c)$. This
framework is called the {\it reduced parameter space}. The existence of
a phase transition for $P<P_c$ is observed.
In Fig. \ref{fig:gt4y8} we considered two cases of the behavior of the EEH-AdS BH,
that are two isobaric lines, one for $P=0.4\, P_c$ and the second one for $P=0.8\, P_c$. It
is observed that the isobaric curve in the $T$-$S$ diagrams can be divided into three
branches in the direction of increasing entropy: {\it small} (coral and purple),
{\it intermediate} (dashed green) and {\it large} (blue), which can help us to
identify the phase transitions.  When two phases coexist, the {\it Maxwell's equal
area law} is fulfilled; i.e., the areas above and below a line of constant
temperature (pressure) through a $T$-$S$ ($P$-$V$) diagram are equal. That is satisfied
by the $AC$, $BD$ and $AB$ lines in Fig. \ref{fig:gt4y8}.
 
The phase transition, when the size of the black hole goes from a certain volume,
$v_1$ to a a bigger one, $v_2$, occurs at the same Gibbs free energy and the same
temperature (or equivalently, the same pressure); i.e., at the phase transition
it must be fulfilled that
\begin{equation}
G(v_1)=G(v_2), \quad T(v_1)=T(v_2).\label{g1=g2}
\end{equation}

For $P=0.4 P_c$ this transition can happen at two temperatures, $T\approx 0.7547
\;T_c$ corresponding to the $AC$ line and $T \approx 0.6823 \;T_c$ for the $BD$
line, as we can see in the upper Fig. \ref{fig:gt4y8}. 
From the diagrams for the Gibbs free energy we can also observe a kind of reentrant phase transition, which means that the final state of a system is macroscopically similar to the initial state.  Starting at the vertical blue line (large BH), the curve reaches the maximum of $G$ and passes to the small BH region (through the intermediate BH), to finish crossing the initial blue line. In other words, the system describes a large-small-large BH reentrant phase transition. More details about reentrant phase transition can be consulted in \cite{Kubiznak2017} and \cite{FrassinoKM2014}.

\begin{figure}[H]
\centering
\includegraphics[scale=0.6]{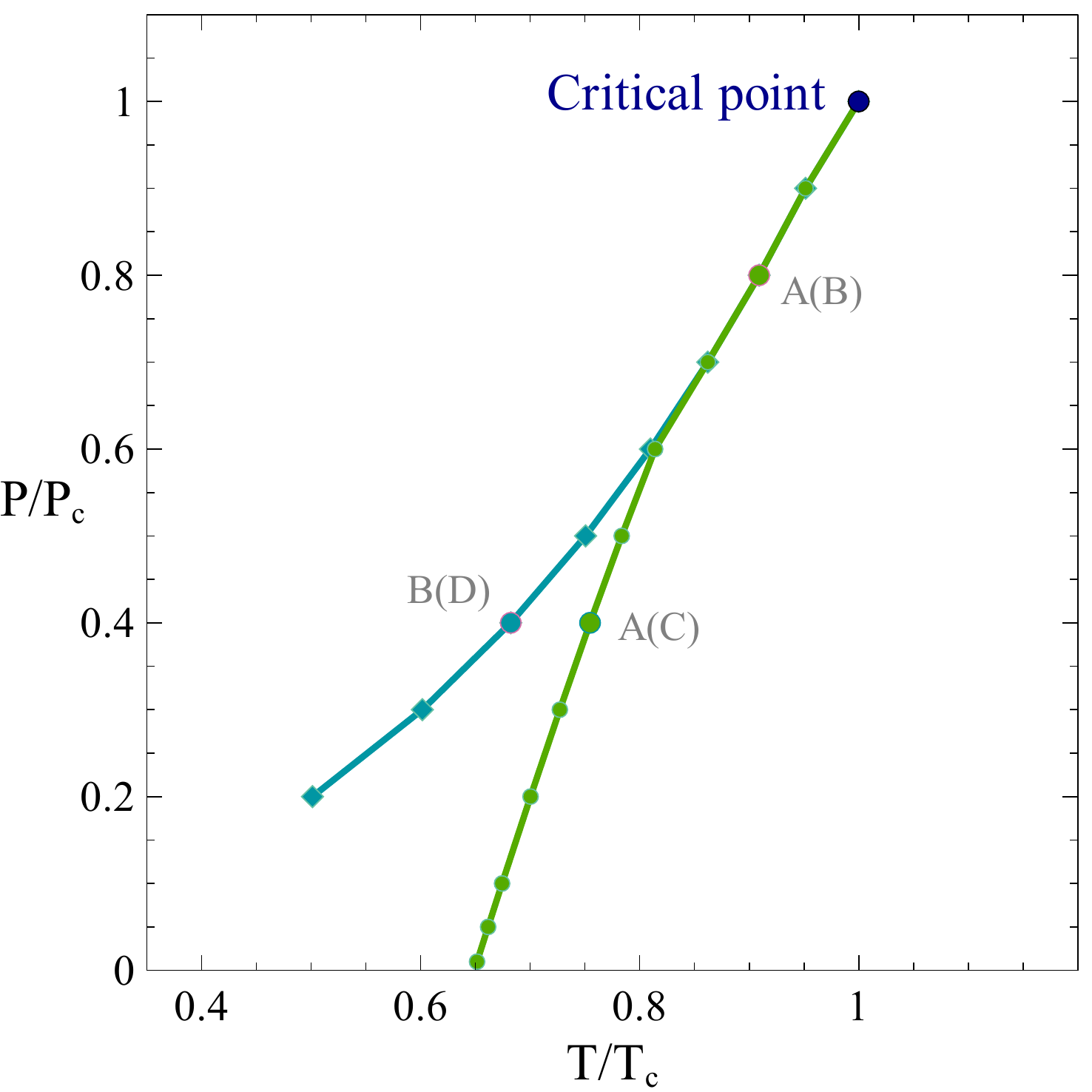}
\caption{Coexistence $P-T$ curve for the EEH-AdS BH. It was numerically determined from the conditions 
in Eq. (\ref{g1=g2}), 
$G(v_1)=G(v_2), T(v_1)= T(v_2) $; from $P\in (0.2, 0.6)P_c$ there are two
different temperatures at which the transition occurs.   
In this plot we fixed  $Q=1$ and $a=1$.}
\label{fig:Coex}
\end{figure}

For the case $P= 0.8 P_c$ the
behavior is a little bit simpler, because we only observe the phase transition for
$T \approx 0.9091\; T_c$, corresponding to the $AB$ line in Fig. \ref{fig:gt4y8} (bottom).
For the black hole under study there is no analytic expression for the coexistence
$P$-$T$ curve; however, in Fig. \ref{fig:Coex}, we plot the numerical values that
satisfy (\ref{g1=g2}). Notice that from $P=0.2\,P_c$ to  $~0.6\,P_c$ there are two
different temperatures at which the transition occurs, resulting in the
coexistence curve separating into two. The region between these two lines can be
identified as an intermediate black hole region. The marked points $A(C),\, B(D)$ and
$A(B)$ correspond to the phase transitions showed in Fig. \ref{fig:gt4y8}.


\subsection{Critical exponents for the EEH-AdS BH.}
Now, we calculate the critical exponents, $\alpha, \beta, \gamma, \delta$, that
govern the behavior of the physical properties near  the critical point
for the EEH-AdS BH.
To calculate them, we are going to make a deduction of the law of corresponding
states using the frame of reduced variables, $p,\tau$ and $\nu$ that are defined by
\begin{equation}
    p\equiv\frac{P}{P_c},\hspace{1cm} \tau\equiv \frac{T}{T_c}, \hspace{1cm}
\nu\equiv\frac{v}{v_c}. \label{reduced}
\end{equation}
Such that, the equation of state (\ref{p(v)}), takes the form

\begin{equation}
p=\frac{1}{\rho_c}\frac{\tau}{\nu} +\frac{1}{\pi P_c
v_c^2}\left(-\frac{1}{2\nu^2}+\frac{2Q^2}{v_c \nu^4}-\frac{8aQ^4}{v_c^6
\nu^8}\right).
\end{equation}

Let us assume that the previous equation can be expressed as
\begin{equation}
    p(\tau,\nu)= \frac{1}{\rho_c}\frac{\tau}{\nu} + h(\nu), \label{p1}
\end{equation}
where $\rho_c$ is the critical ratio. The Taylor expansion in the vicinity of the
critical point,  $\tau=1$, $\nu=1$ is written as
\begin{equation}
    p\approx
1+\frac{1}{6}\left(-\frac{6}{\rho_c}+h^{(3)}(1)\right)(\nu-1)^3+\frac{1}{\rho_c}(\tau-1)-\frac{1}{\rho_c}(\nu-1)(\tau-1)+\frac{1}{\rho_c}(\nu-1)^2(\tau-1),
\end{equation}
where we have used that, $\partial_\tau p (1,1)=\partial_\tau p(1,1)=0$, from the
definition of the critical point.
Defining 
\begin{equation}
    t\equiv \tau -1, \hspace{1cm} w\equiv \nu-1,\label{t,w}
\end{equation}
the law of corresponding states takes the form
\begin{equation}
    p(t,w) = 1+\frac{1}{\rho_c} t-\frac{1}{\rho_c}wt-Cw^3+\mathcal{O}(t w^2,w^4),
\label{p(t,w)}
\end{equation}
where we have defined $C=\left(\frac{1}{\rho_c}-\frac{h^{(3)}(1)}{6} \right)$, using Eqs. (\ref{Crit}) and (\ref{rho_app})
we can approximate it, getting
\begin{equation}
C\approx\frac{4}{3}-0.002  \frac{(129.019 - 10.034 Q^2 +  Q^6)a}{Q^4}.
\end{equation}
  
The next step is to determine the relation between  the volume of the BH in
the phase transition $w_a$ and  $w_b$. From the Maxwell's equal area law we know
that

\begin{eqnarray}
0&=&\int_{a}^{b} v dP=v_c P_c \int_{w_a}^{w_b}w\left(\frac{1}{\rho_c}t+3Cw^2
\right) dw\nonumber\\ \vspace{.5cm}
&\Rightarrow &\hspace{.2cm}
w_a^2\left(\frac{t}{2\rho_c}+\frac{3C}{4} w_a^2\right)=
w_b^2\left(\frac{t}{2\rho_c}+\frac{3C}{4} w_b^2\right),\nonumber
\end{eqnarray}
i.e. $w_a= - w_b$, because for a given isotherm 
$dp=\frac{\partial p}{\partial w} dw = -\left(\frac{1}{\rho_c}t+3Cw^2 \right) dw $
and in the coexistence of the phase transition  $p_a=p_b$. Thus
\begin{eqnarray}
\frac{1}{\rho_c}w_a\,t+Cw_a^3&=&\frac{1}{\rho_c}w_b\,t+Cw_b^3\nonumber\\
\Rightarrow\hspace{.2cm} w_a=\sqrt{-\frac{1}{\rho_c}\frac{t}{C}}\,&=&\,
\frac{v_a}{v_{ac}}-1.\label{w_s}
\end{eqnarray}
It only remains to read the values of the critical exponents $\alpha, \beta, \gamma$ and
$\delta$ from the previous expressions.

\begin{itemize}
\item $\alpha\rightarrow$ Is related with the behavior of the specific heat at
constant volume $C_v$.
\begin{equation}
C_v=T\left(\frac{d S}{dT}\right)_v \propto |t|^{-\alpha}.\label{alpha}
\end{equation}
    
Since the entropy $S$ has no dependence on $T$, it follows that
\begin{equation}
C_v=0\hspace{.2cm}\rightarrow\hspace{.2cm} \alpha=0.\nonumber
\end{equation}
    
\item $\beta\rightarrow$ Is related to the order parameter $\eta$, namely with
the change of volume at the phase transition for a given isotherm,
    
\begin{eqnarray}
\eta = v_g-v_l = v_c(w_s-w_l)&\propto& |t|^\beta.\nonumber\\
w_s&\propto& |t|^{1/2}\hspace{.2cm} \rightarrow\hspace{.2cm} \beta=1/2.
\end{eqnarray}
    
\item $\gamma\rightarrow$ Regulates the behavior of the isothermal
compressibility $\kappa_T$
\begin{eqnarray}
        \kappa_T = -\frac{1}{v} \left(\frac{\partial v}{\partial P}\right)_T=
\frac{1}{P_c(w+1)}\left(\frac{\partial p}{\partial w}\right)^{-1} &\propto&
|t|^{-\gamma}\nonumber\\
        \kappa_T\approx\frac{1}{P_c A t} &\propto&|t|^{-1} \hspace{.2cm}
\rightarrow\hspace{.2cm} \gamma= 1.
\end{eqnarray}
\item $\delta\rightarrow$ In a critical isotherm, $T=T_c$ or equivalently $t=0$.
\begin{eqnarray}
        |P-P_c|&\propto& |v-v_c|^\delta\nonumber  \\
        |p-1|&\propto& |w|^\delta\nonumber\\
        p-1 &=&C w^3 \hspace{.2cm} \rightarrow\hspace{.2cm} \delta=3.
\end{eqnarray}
\end{itemize}
 
The obtained critical exponents  are:
$\alpha=0, \beta= 1/2, \gamma=1, \delta=3,$. Those are the same as the ones in the standard mean field theory of charged AdS black holes \cite{Kubiznak}, thus exhibiting that the universality of
such kind of phase transitions  includes the EEH-AdS BH.

\section{Conclusions}

In this paper we generalized the static spherically symmetric solution to the 
Euler-Heisenberg electrodynamics coupled to Einstein gravity, by including the 
cosmological constant $\Lambda$.  The introduction of $\Lambda$ allows the
thermodynamic analysis of the EEH-AdS BH, under the interpretation of $\Lambda$ as a
thermodynamic pressure.

From the analysis of the effective potential for test particles, as a result of 
introducing $\Lambda$,  the possibility exists of stable and unstable circular orbits for test particles
both massless and massive.  Then we focus on its
thermodynamic properties. Based on dimensional analysis we derived the  Smarr formula for the
mass (enthalpy in the canonical ensemble) of the EEH-AdS BH in terms of its charge $Q$, the anti--de Sitter parameter $\Lambda$,
and the EH parameter $a$; the latter we promote to thermodynamic variable, interpreting $a$ as the vacuum polarization, being its conjugate variable $\mathcal{A} = \left( \partial M / \partial a \right)$
that has units of electric field. The proposed Smarr formula turns out to be in agreement with the first law of BH thermodynamics. We also present the equation of state and the $P$-$v$ diagrams showing
the phase transitions small/large BH, that occur for $P <P_c $ and $T< T_c$.
One of the effects of the EH parameter is to 
increase the critical values of the temperature $T_c$ and the pressure $P_c$.
Moreover, the analysis of the critical values of the pressure, temperature and
volume  shows that the system presents two behaviors: one close
to the Maxwell case (linear electrodynamics) while the other one is manifestly
nonlinear electromagnetic. Besides $P_{\rm crit} v_{\rm crit}/ T_{\rm crit} = 3/8 + \mathcal{O}(a) \times 10^{-4} $.

The behavior of the specific heat at fixed charge $C_Q$ points to a phase transition that we confirm from
the characteristic swallowtail behavior of the Gibbs free energy that leads one to conclude the
existence of first order phase transitions for $P<P_c$,
fulfilling the Maxwell's equal area law  in the $T$-$S$ diagrams.
We also display the coexistence curve $P$-$T$, which allows us to see clearly the 
{\it small/intermediate/large} black hole phase transitions. Moreover, 
we constructed the law of corresponding states in the reduced parameter space near to
the critical point, leading to the determination of the critical exponents, which
are exactly the same for the Van der Waals fluid,  thus confirming for the 
EEH-AdS BH the universality of such kinds of phase transitions.
One relevant feature of this system is a split in the small BH region, 
which results in a second (reentrant) phase transition for certain values of the pressure for
the EEH-AdS BH.

\vspace{0.5cm}
\textbf{Acknowledgments}: The work of DM has been sponsored by CONACYT-Mexico through the PhD. Scholarship No. 434578.  N. B. acknowledges partial financial support from CONACYT-Mexico through the Project No. 284489.


\end{document}